\documentclass[nofootinbib, aps,twocolumn,showpacs,preprintnumbers,amssymb,superscriptaddress,10pt]{revtex4-2}
\pdfoutput=1

\usepackage{amsmath,amssymb,amsfonts,graphicx,epsfig}
\usepackage[noconfig]{refstyle}
\usepackage[hyperfootnotes=true, linktocpage=true, citecolor=blue, linkcolor=blue, urlcolor=Maroon, filecolor=Maroon]{hyperref}

\newcommand{\ii}{\mathrm{i}}

\let\eqref=\relax
\newref{eq}{name={},Name={Eq.~},names={eqs.~},Names={Eqs.~},rngtxt={-},refcmd=(\ref{#1})}
\newref{tab}{name={Table~},Name={Table~},names={tables~},Names={Tables~}}
\newref{sec}{name={Section~},Name={Section~},names={sections~},Names={Sections~}}
\newref{fig}{name={Figure~},Name={Figure~},names={figures~},Names={Figures~}}
\numberwithin{equation}{section}

\newcommand{\comment}[1]{}

\usepackage{multirow}
\usepackage{subcaption}
\usepackage{float}

\begin{document}

\title{Machine-Learning the Classification of Spacetimes}

\author{Yang-Hui He}
\affiliation{London Institute, Royal Institution of GB, 21 Albemarle St., London, W1S 4BS}
\affiliation{Merton College, University of Oxford, OX1 4JD, UK}
\affiliation{Department of Mathematics, City, University of London, London EC1V0HB, UK}
\affiliation{School of Physics, NanKai University, Tianjin, 300071, P.R. China}
\email[]{hey@maths.ox.ac.uk} 

\author{Juan Manuel P\'erez Ipi\~na}
\affiliation{Mathematical Institute, University of Oxford,\\
Andrew Wiles Building, Radcliffe Observatory Quarter,\\
Woodstock Road, Oxford, OX2 6GG, U.K.}
\email[]{Juan.PerezIpina@maths.ox.ac.uk}

\begin{abstract}\noindent
On the long-established classification problems in general relativity
we take a novel perspective by adopting fruitful techniques from machine learning and modern data-science.
In particular, we model Petrov's classification of spacetimes, and show that a feed-forward neural network can achieve high degree of success. We also show how data visualization techniques with dimensionality reduction can help analyze the underlying patterns in the structure of the different types of spacetimes.
\end{abstract}


\maketitle

\section{Introduction \& Summary}
What are the possible structures of spacetime?
This is surely one of the most important questions in theoretical physics.
Classification problems in general relativity have been an active field since the very beginning and have more recently been a focus of computer algebra systems \cite{Stephani:2003tm,MacCallum:2018csx}. 
Fully classifying and comparing Riemannian manifolds can be achieved through the Cartan-Karlhede algorithm \cite{Karlhede1980}. The first step in this algorithm is to determine the Petrov \cite{Petrov:1954} and Segre \cite{Segre1884} types of the spacetime \cite{Stephani:2003tm,2000CQGra..17.2885P}. These methods analyze algebraic symmetries of the Weyl and Ricci tensor, respectively, and involve detailed study of roots and multiplicities of certain quartic equations. In particular, Petrov's classification of the Weyl tensor has been an integral part of the study of exact solutions to the Einstein equations. Here, we will illustrate a new computational approach that can be used in the Petrov classification problem, which can then be extended for a full classification of gravitational solutions.

Since the recent introduction of machine-learning and related techniques of modern data science, to study the string theory landscape \cite{He:2017aed,He:2017set,Krefl:2017yox,Carifio:2017bov,Ruehle:2017mzq}, and more generally the vast landscape of pure mathematics \cite{He:2019nzx,Alessandretti:2019jbs,He:2020fdg,He:2021oav,davies2021advancing}, it is natural to address our present problem of spacetime classification under the auspices of this programme.
The reader is also referred to the pedagogical introduction of machine-learning in theoretical physics and mathematics by \cite{He:2018jtw,Ruehle:2020jrk} as well as references therein.
Furthermore, detection of symmetries in physical systems relevant to our context, using machine-learning, is also discussed in 
\cite{Krippendorf:2020gny,Krishnan:2020sfg,Chen:2020dxg,Liu:2021azq,Alexander:2021rch,Altman:2021pyc,Gao:2021xbs}.

In this letter, we apply some of these machine-learning (ML) techniques to Petrov's classification of spacetimes. Since the original formulation of the problem, many algorithms have been proposed to model the classification (see, for example, \cite{dInverno1971CLASSIFICATIONOT,LM1988,1991GReGr..23.1023A,2000CQGra..17.2885P,Zakhary:2003}). These usually reduce the problem to finding the roots and multiplicites of a quartic equation where the parameters are a set of five complex Weyl scalars $\Psi_i$ ($i=0,...,4$) in the Newman-Penrose formalism \cite{doi:10.1063/1.1724257}. These Weyl scalars can easily be computed for any spacetime and the relations between the nonvanishing $\Psi_i$ determine the Petrov type of the manifold. Here we take this approach for building a supervised learning problem fit for ML tools. 

In Section \ref{sec:data} we give an overview of the problem and show how to represent the spacetime data in an expedient manner.
We artificially generate numerical data to train and validate various ML classifiers. Specifically, we start by building different datasets of Weyl scalars $\{\Psi_0,\Psi_1,\Psi_2,\Psi_3,\Psi_4\}$, with randomly generated entries, and then manually labeling each data point with its corresponding Petrov type. These datasets are later used in Section \ref{sec:NN} to train several ML classifiers to see how well they learn and compare. 
We find that feed-forward neural networks (NN) are the most accurate classifiers for this problem, obtaining very high precision in only a handful of epochs. Moreover, in Section \ref{sec:PCA}, we use other data science techniques, like Principal Component Analysis (PCA), to further study latent patterns in the data, that give rise to the Petrov classification. We show how data visualization tools can illustrate the intrinsic differences between spacetimes of distinct Petrov type. Finally, we discuss the results and future applications of this programme in Section \ref{sec:end}.

\section{The Petrov Classification} \label{sec:data}
\begin{figure}
\centering
\begin{subfigure}{.25\textwidth}
  \centering
  \includegraphics[width=.8\textwidth]{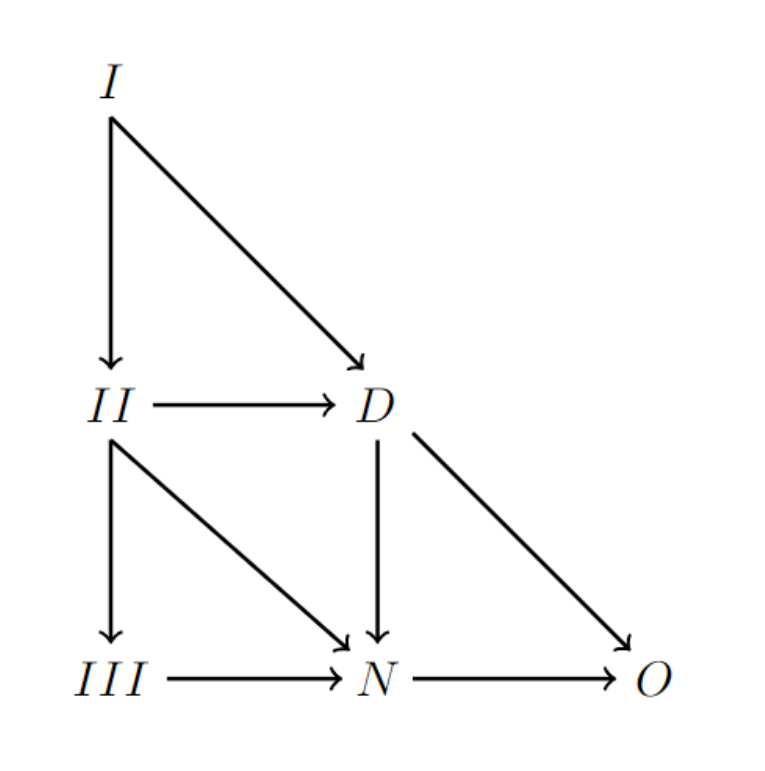}
  \label{fig:PetrovMults1}
\end{subfigure}%
\begin{subfigure}{.25\textwidth}
  \centering
  \includegraphics[width=.9\textwidth]{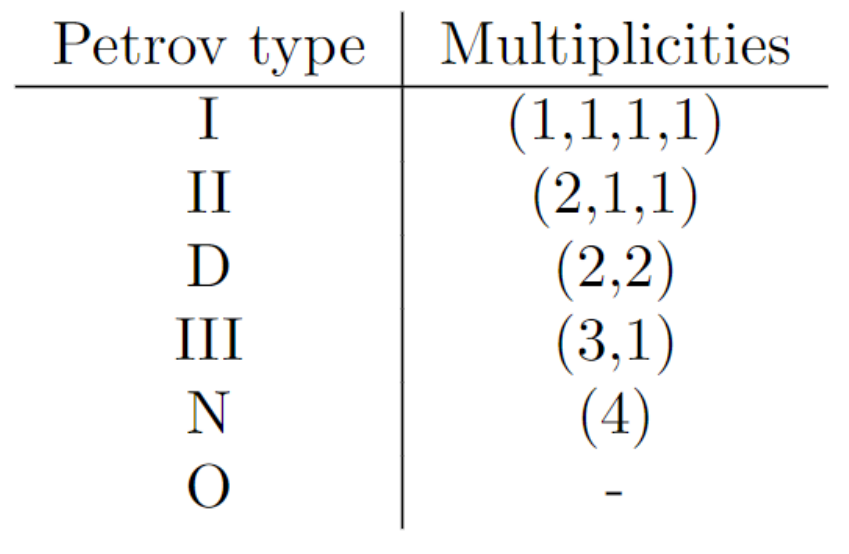}
  \label{fig:PetrovMults2}
\end{subfigure}
\caption{Classification of the Petrov type according to the number and multiplicity of principal null directions (with the arrows denoting possible degenerations of one Petrov type into another). 
Type O corresponds to the vanishing of the Weyl tensor and so does not single out any principal null directions.}
\label{fig:PetrovMults}
\end{figure}

Petrov's classification of the algebraic symmetries of the Weyl tensor can be formulated as an eigenvalue problem for the Weyl tensor evaluated at some spacetime event. Alternatively, one can see it as a characterization of the Weyl tensor in terms of the principal null directions (p.n.d.) at that event \cite{PENROSE1960171} (see the Appendix for details into this approach). Depending on the amount and multiplicity of the p.n.d.'s we can classify spacetimes in $6$ distinct types: $I,II,III,D,N,O$. The classification can be seen in Figure \ref{fig:PetrovMults}.

As is shown in the Appendix, one can see that the determination of principal null directions is equivalent to solving the following quartic equation for $z$: 
\begin{equation} \label{quartic}
\Psi_0-4z \Psi_1 +6z^2 \Psi_2-4z^3 \Psi_3 +z^4 \Psi_4 = 0\,,
\end{equation}
where $\Psi_i$ ($i=0,1,2,3,4$) are the five complex Weyl scalars in the Newman-Penrose formalism, and are defined in (\ref{PsiDef}).

\paragraph{The $n=32$ Cases: }
The vanishing of one or more of these $\Psi_i$ simplifies (\ref{quartic}), and this has been a staple of most attempts to determine the Petrov type. This has been taken into account in many of the aforementioned algorithms where, starting with the work of \cite{LM1988}, a parameter $n$ was introduced to distinguish the $32$ possible combinations of vanishing/non-vanishing Weyl scalars. Each of these $32$ classes might have one or more Petrov types assigned to it. For a detailed list of the classes and their Petrov types, see Table \ref{table32} (where we've ordered the cases according to the number of vanishing Weyl scalars, and not on the value of $n$ from \cite{LM1988}).

As can be seen from Table \ref{table32}, for the cases with 3 or more vanishing Weyl scalars the Petrov type can be immediately determined; this is not the case for the rest. When working with an arbitrary null tetrad, the Weyl vector $\{ \Psi_i\}$ might be arbitrarily hard and it takes more work to determine the Petrov type. Of course, the Petrov classification is coordinate independent and specifically, does not depend on the choice of tetrad, as long as that frame is not singular \cite{Tanatarov:2012gf}. To distinguish between the possible types at the bottom half of the table, there have been many analytical results as in \cite{LM1988,2000CQGra..17.2885P} (building polynomials out of the remaining non-vanishing Weyl scalars). Since we want our classifier to handle completely general data (and work in any basis), we want to train in all possible cases of Table \ref{table32}.

\begin{table}
\centering
\resizebox{\columnwidth}{!}{
\begin{tabular}{ |c|c|c|c| } 
\hline
Number of zeros & Form & Petrov type \\
\hline
5 & 00000 &  O \\ 
\hline
\multirow{5}{*}{4} & N0000 & \multirow{2}{*}{N} \\ 
& 0000N & \\ \cline{2-3}
& 00N00 &  D \\ \cline{2-3}
& 0N000 & \multirow{2}{*}{III}  \\ 
& 000N0 &  \\ 
\hline
\multirow{10}{*}{3} & NN000 & \multirow{2}{*}{III} \\ 
& 000NN & \\ \cline{2-3}
& 00N0N &  \multirow{4}{*}{II} \\ 
& N0N00 &  \\ 
& 00NN0 & \\ 
& 0NN00 &  \\ \cline{2-3}
& 0N00N & \multirow{4}{*}{I}  \\ 
& N00N0 &  \\ 
& 0N0N0 & \\ 
& N000N &  \\ 
\hline
\multirow{10}{*}{2} & 00NNN & \multirow{2}{*}{II or D} \\ 
& NNN00 & \\ \cline{2-3}
& N0N0N &  I or D \\ \cline{2-3}
& 0N0NN &  \multirow{7}{*}{I or II}\\ 
& NN0N0 & \\ 
& N00NN &  \\ 
& NN00N &   \\ 
& 0NN0N &  \\ 
& N0NN0 & \\ 
& 0NNN0 &  \\ 
\hline
\multirow{5}{*}{1} & 0NNNN & \multirow{4}{*}{I, II or III} \\ 
& NNNN0 & \\ 
& N0NNN & \\ 
& NNN0N &  \\ \cline{2-3}
& NN0NN &  I, II, III or D\\ 
\hline
0 & NNNNN & I, II, III, D or N \\
\hline
\end{tabular}
}
\caption{Determination of the Petrov type according to the vanishing of the five Weyl scalars $\{\Psi_0,\Psi_1,\Psi_2,\Psi_3,\Psi_4\}$. ``Form'' refers to the vanishing of the five quantities $\Psi_i$:
N signifies a non-vanishing entry and 0, a vanishing one. 
This table was based on the one at \cite{2000CQGra..17.2885P}, where we also corrected some typos.}
\label{table32}
\end{table}

\begin{figure*}
\centering
\includegraphics[width=\textwidth]{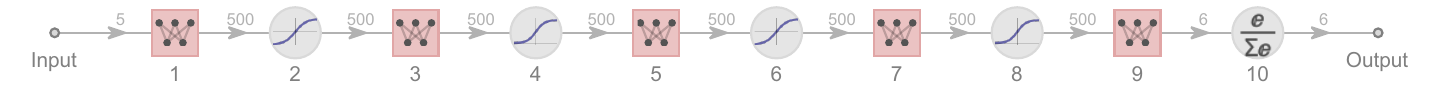}
\caption{Architecture of the five-layer neural network. The hidden layers are alternating between $\tanh$ and sigmoid activated; the last layer corresponds to the softmax activation function. The hidden layers contain $500$ nodes each, and the softmax has $6$, corresponding to the six output classes.}
\label{fig:NN}
\end{figure*}

\paragraph{Data Generation: }
For this purpose we treat $\{\Psi_0,\Psi_1,\Psi_2,\Psi_3,\Psi_4\}$ as a numerical five-vector, randomly generating the entries for every possible case and subcase in Table \ref{table32}.

We created different databases formed from integer, rational, real or complex entries. The latter two permit the creation of huge datasets uniformly distributed in a specific range (e.g. for the reals $\Psi_i  \in \{ -10,10 \}$). Unfortunately, for the real and complex data points, some subcases where not possible to sample through purely random generation so the analytical results of \cite{LM1988,2000CQGra..17.2885P} were used to generate this remaining data.

Specifically, for the real (or complex) dataset, $10,000$ points were collected from each $n$ except the last case, NNNNN, where $20,000$ points were sampled. This amounts to a total size of $330,000$ data points, of different Petrov types. Notice that by doing this we are taking a different number of data points per Petrov type but this is consistent with how common it is to find each type. For example, for the real dataset the resulting tally of points per type can be seen in Table \ref{tab:tally}. These vectors were then labeled by their corresponding Petrov type, through the implementation of the \cite{dInverno1971CLASSIFICATIONOT} algorithm in Mathematica \cite{Mathematica}.

\begin{table}[H]
    \centering
    \resizebox{\columnwidth}{!}{
    \begin{tabular}{|c|c|c|c|c|c|c|}
    \hline
        I & II & III & D & N & O & Total\\
        \hline
        126,000 & 90,500 & 55,500 & 24,500 & 23,500 & 10,000 & 330,000\\
        \hline
    \end{tabular}
    }
    \caption{Tally of data points per Petrov type for the dataset of real entries. The distribution of points per class is not homogeneous and this has to be taken into account when judging the efficiency of a classifier.}
    \label{tab:tally}
\end{table}

\section{Building a classifier} \label{sec:NN}
For our supervized ML paradigm, the above dataset was randomly split in three groups: $70\%$ for the training set, $15\%$ for validation and $15\%$ for testing. The first two are used to train the classifiers, while the testing set is used to evaluate the performance in never before seen data. Many different types of classifiers were trained and tested, including: decision trees (boosted), random forests, nearest neighbours, and more. While these methods achieved reasonable accuracies, the best results were obtained using feed-forward neural networks (NN), which we detail shortly.

\begin{figure}[H]
\begin{subfigure}{.5\textwidth}
  \centering
  \includegraphics[width=\textwidth]{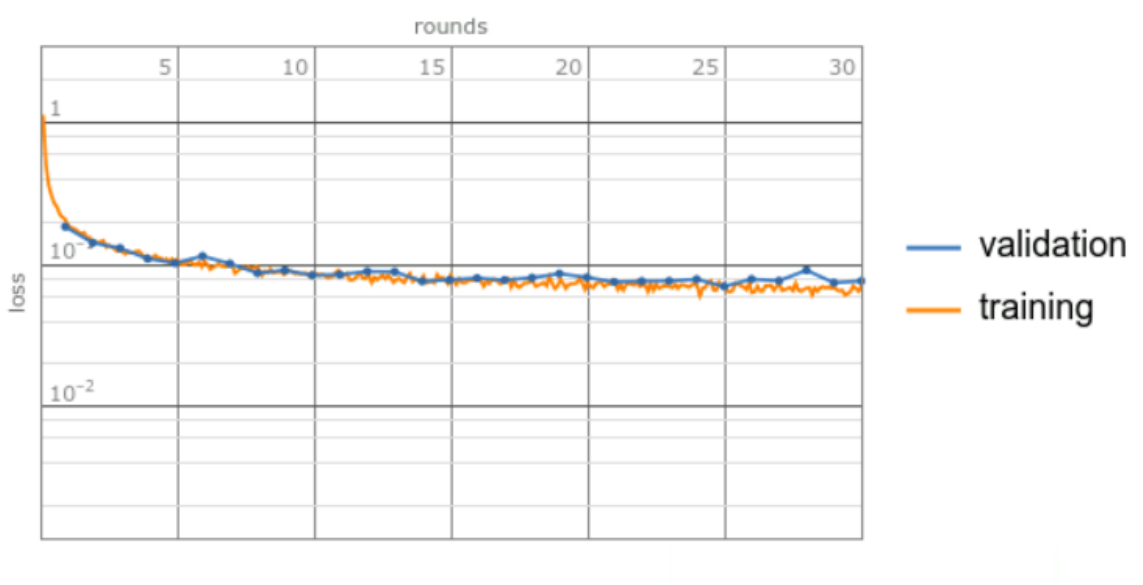}
  \label{fig:loss}
\end{subfigure}
\begin{subfigure}{.5\textwidth}
  \centering
  \includegraphics[width=\textwidth]{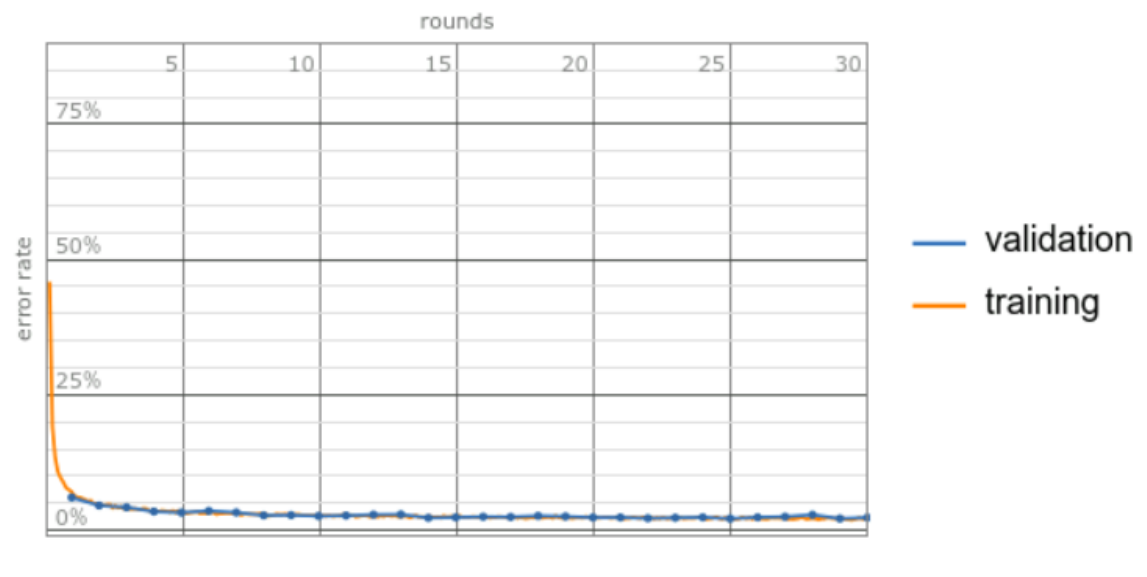}
  \label{fig:errorrate}
\end{subfigure}
\caption{Loss-function (above) and error rate (below) for the training of the neural network, plotted against the number of epochs or rounds.}
\label{fig:NNplots}
\end{figure}

\begin{figure*}
\begin{subfigure}{.19\textwidth}
  \centering
  \includegraphics[width=.9\textwidth]{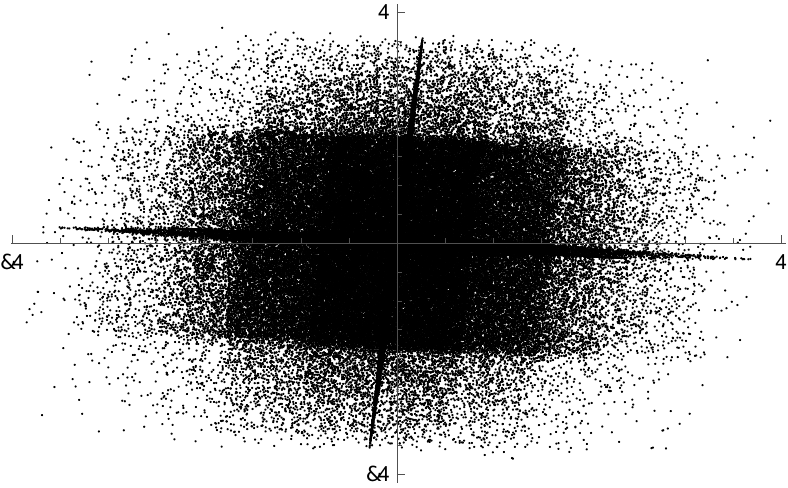}
  \caption{Type I}
  \label{fig:PCAmultiI}
\end{subfigure}
\begin{subfigure}{.19\textwidth}
  \centering
  \includegraphics[width=.9\textwidth]{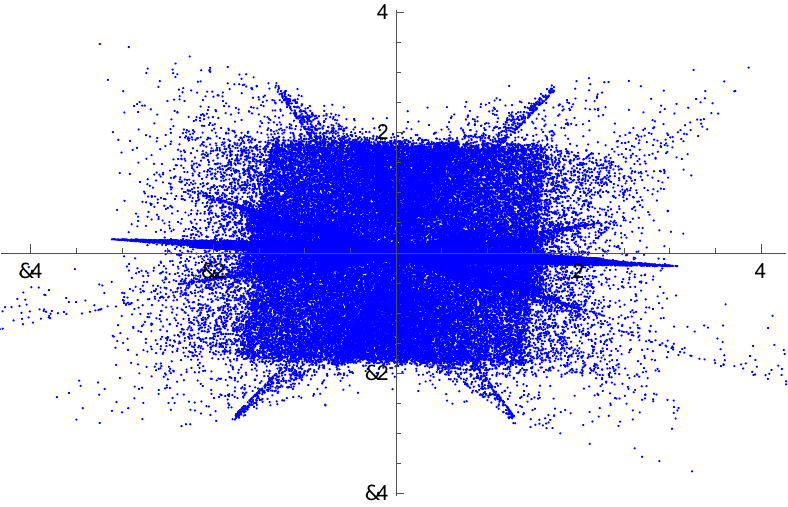}
  \caption{Type II}
  \label{fig:PCAmultiII}
\end{subfigure}
\begin{subfigure}{.19\textwidth}
  \centering
  \includegraphics[width=.9\textwidth]{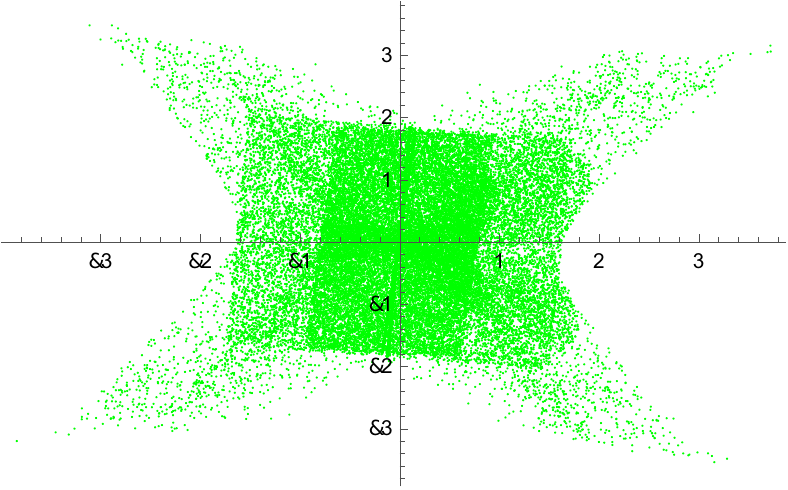}
  \caption{Type III}
  \label{fig:PCAmultiIII}
\end{subfigure}
\begin{subfigure}{.19\textwidth}
  \centering
  \includegraphics[width=.9\textwidth]{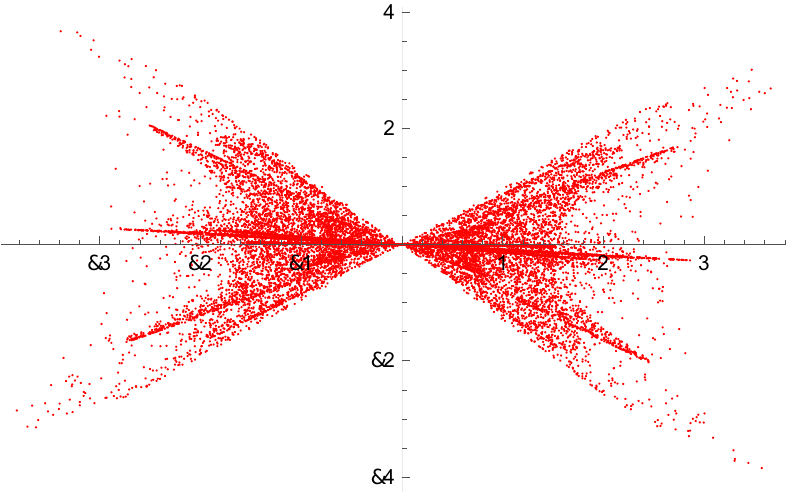}
  \caption{Type D}
  \label{fig:PCAmultiD}
\end{subfigure}
\begin{subfigure}{.19\textwidth}
  \centering
  \includegraphics[width=.9\textwidth]{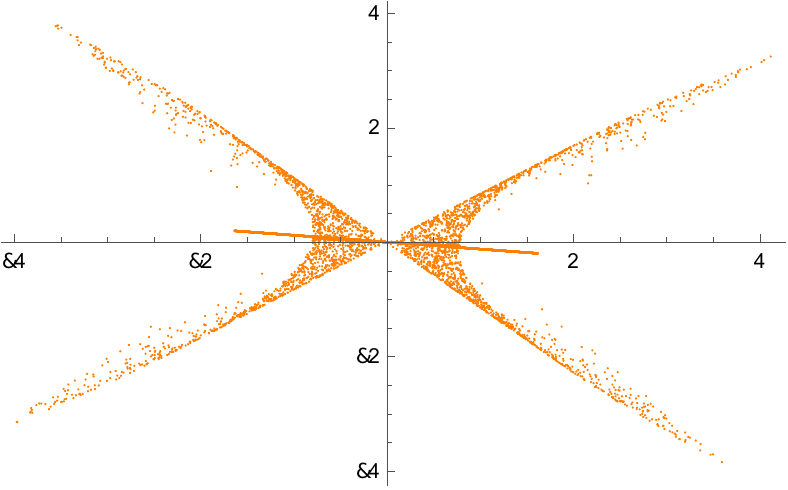}
  \caption{Type N}
  \label{fig:PCAmultiN}
\end{subfigure}
\caption{Visual representation of the different Petrov types. The data was dimensionally reduced using Principal Component Analysis (PCA) to observe the directions with the highest variance.}
\label{fig:PCAmulti}
\end{figure*}

The non-linearity in a NN model is obtained through the choice of activation functions. For this problem we found the highest accuracy in the use of hyperbolic tangents and logistic sigmoids. While the problem can be modelled using a single hidden layer, we found higher accuracy in fewer epochs when using multiple hidden layers. In Figure \ref{fig:NN} we show the architecture of our NN that combines both activation functions in multiple alternating hidden layers. It takes as input the five-dimensional\footnote{Here and in the following we will use the dataset built from real entries for $\Psi_i$. An analog analysis was produced for the complex dataset, where the input vector is ten-dimensional, after splitting in real and imaginary parts. Similar results in accuracy and confidence were found for the complex dataset.} vectors of Weyl numbers, then goes through four hidden layers of $500$ nodes each, with alternating activation functions: $\tanh$ and sigmoids. The specific numbers of nodes and hidden layers were also found to produce the highest accuracy results, but by no means do we claim this to be the most efficient configuration possible. Different choices of these hyperparameters (or other variables such as the optimizer, the label encoding or the learning rate) represent possible directions of improvements on this neural network. Finally, since we have here a multi-class classification problem, the last layer is a softmax layer, with 6 nodes for the 6 different classes. 

As mentioned above, the NN from Figure \ref{fig:NN} was trained and optimized using the training and validation sets, and the testing set was used to determine its accuracy. The network was trained for 30 epochs, using a learning rate of $10^{-3}$, and the ADAM optimizer \cite{kingma2017adam}. In Figure \ref{fig:NNplots} one can see the steady decrease of the loss function and error rate, as the number of training rounds increases. We define accuracy as percentage agreement of predicted versus actual values. However, when dealing with imbalanced multi-class classification problems accuracy is not the most useful evaluation metric. To take these differences into account we define confidence then through the use of Matthew's Correlation Coefficient (MCC) $\varphi$ generalized to the multi-class case \footnote{
Alternatively, we can also compute the $F_1$-score for each class and then the weighted $F_1$-score for the whole dataset.
In our calculations, we find the $F_1$-score to be $0.979$. 
}.
In all, we achieved an accuracy of $0.979$, confidence of $0.973$, and a final loss of $6.61\times 10^{-2}$.

We can plot the confusion matrix to see the successes and mistakes for each class. 
This is a $6 \times 6$ integer matrix of the actual numbers in the Petrov class of ($O, I, II, III, D, N$) versus the numbers as predicted by the ML classifier.
\begin{figure}
\centering
\includegraphics[width=0.45\textwidth]{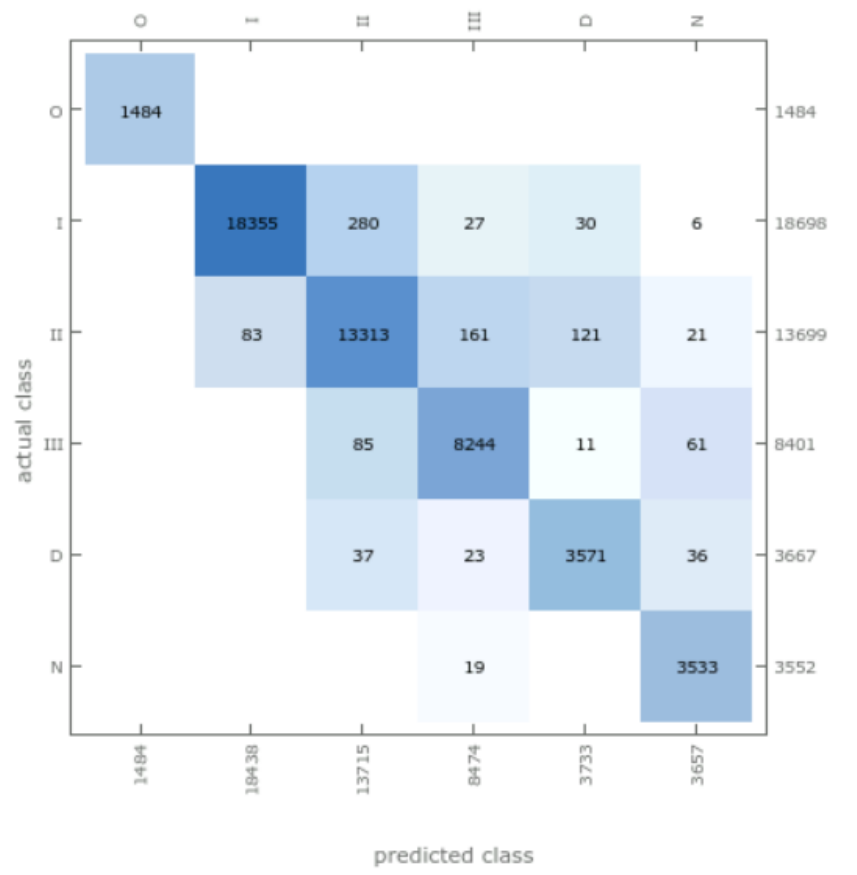}
\caption{A plot of the Confusion Matrix by our NN classifier; we can see that it is heavily diagonal, signifying that the classification into the 6 Petrov types is extremely accurate.}
\label{fig:confumatrix}
\end{figure}

\section{Data visualization} \label{sec:PCA}
Having successfully trained a neural network (as well as other classifiers) in learning the Weyl data, it is also interesting to see how our data looks, and what patterns we directly observe.
This is very much in the spirit of conjecturing formulation via ML \cite{He:2021oav}, to let ML algorithms detect patterns which might ab initio be hidden.
For this we can follow a standard Principal Component Analysis (PCA) to dimensionally reduce the data to its highest-variance components so we can study the resulting two-dimensional plots. 

We first obtain the principal components from the full unlabelled dataset and then reattach the corresponding labels (color-coding each distinct class for visualization). In Fig. \ref{fig:PCAmulti} we can see the principal component representation for each Petrov type (not type O since it corresponds only to the vector $\{0,0,0,0,0\}$ and therefore has no variation). Note that within the populated areas of the plots, some are more densely populated than others, reflecting the specific data generation procedure of Section \ref{sec:data}. 

One can see how in the most general case, the data is spread out everywhere with no pattern in sight. As we increase the degeneration (that is, we move downward in Figure \ref{fig:PetrovMults}, the data starts settling into definite patterns.

In particular, types D and N have very specific shapes, illustrating the particularity of these cases. One can for instance superpose these figures and see exactly how one Petrov type degenerates into another, but for clarity we do not do this since overlaying will obscure many points in the plot.

\section{Outlook} \label{sec:end}
In this work we have shown how to apply techniques from machine learning and data science in classification problems in general relativity. Taking as an example the Petrov classification of the Weyl tensor, we have adapted the problem to fit into the realm of supervised machine learning. 

That is, our input consisted of randomly generated five-dimensional vectors representing the Weyl scalars $\{\Psi_i\}$ ($i=0,...,4$), labelled with their corresponding Petrov type ($I,II,III,D,N,O$). We generated enough data points to consider all possible cases of non-vanishing Weyl components, as described by Table \ref{table32}, to have a set of base-independent training data. We designed a feed-forward neural network to train on this data and achieved $98\%$ accuracy with a confidence (MCC) of $0.973$. This shows that with a very simple neural network, in only a handful of epochs, one can model the Petrov classification with a high degree of success. We also showed how data visualization and dimensionality reduction can help in analyzing the data itself and the patterns that underlie it. Both these directions can help illuminate the intricacies of the classification of spacetimes, shedding light on problems of numerical relativity or in the general study of solutions to the Einstein equations.

The Petrov classification is only a part of the general programme for classifying and comparing spacetimes. The procedure elucidated on this paper can easily be extended to model the Segre classification of the Ricci tensor, for another part of the puzzle. This then constitutes the first step in having a machine ready setup for the full classification of spacetimes, a machine learning formulation of the Cartan-Karlhede algorithm. With the development of online databases for exact solutions to the Einstein equations, the stage is set for a complete exploration of the power of these techniques in this important field.

\section{Acknowledgements}
YHH would like to thank STFC for grant ST/J00037X/2.
\appendix
\section{The Petrov-Penrose Classification of Spacetimes}
In this appendix we provide the conventions and mathematical background used to define Petrov's classification, basing our analysis on \cite{Stephani:2003tm}.
This also sets the notation for the main text, especially Figure \ref{fig:PetrovMults}.

A complex null tetrad is a choice of two real null vectors $\mathbf{l}, \mathbf{k}$ and two complex conjugate null vectors $\mathbf{m}, \mathbf{\bar{m}}$:
\begin{equation}
{\mathbf{e}_a} = (\mathbf{m}, \mathbf{\bar{m}}, \mathbf{l}, \mathbf{k})\,,
\end{equation}
with the only non-vanishing products
\begin{equation}
k^a l_a = -1\,, \qquad m^a \bar{m}_a = 1\,,
\end{equation}
and where the metric in this basis reads
\begin{equation}
g_{ab} = 2 m_{(a} \bar{m}_{b)} -2 k_{(a} l_{b)}\,.
\end{equation}
From this tetrad we can build a basis of bivectors with components
\begin{align} 
\nonumber
U_{ab} &= - l_a \bar{m}_b + l_b \bar{m}_a \,, 
 \\ \nonumber
V_{ab} &= k_a m_b - k_b m_a \,, 
 \\ 
 \label{bivectors}
W_{ab} &= m_a \bar{m}_b - m_b \bar{m}_a - k_a l_b + k_b l_a\, ,
\end{align}
that will be useful in the following.

We remember that the Weyl tensor is the trace-free part of the curvature tensor, given by
\begin{align} \label{defweyl}
C_{abcd} = & R_{abcd} +\frac{1}{2} (R_{bc} g_{ad} + R_{ad} g_{bc} - R_{bd} g_{ac} - R_{ac} g_{bd}) \nonumber\\
 & + \frac{1}{6} R (g_{ac} g_{bd} - g_{ad} g_{bc})\,.
\end{align}
This tensor has the same symmetries as the Riemann curvature, with the added property of tracelessness. In general, it has ten independent components.

For the classification it is useful to define the complex tensor
\begin{equation}
C^{*}_{abcd} \equiv C_{abcd} + \ii C^{\char`~}_{abcd}
\end{equation}
where
\begin{equation}
C^{\char`~}_{abcd} \equiv \frac{1}{2} \varepsilon_{cdef}C_{ab}^{ef} \ . 
\end{equation}
Now we can expand $C^{*}_{abcd}$ in the basis (\ref{bivectors}) as
\begin{align}
\frac{1}{2}C^{*}_{abcd} = & \Psi_0 U_{ab} U_{cd} + \Psi_1 (U_{ab} W_{cd} + W_{ab} U_{cd}) \nonumber\\
& + \Psi_2 (V_{ab} U_{cd} + U_{ab} V_{cd} + W_{ab} W_{cd}) \\
& + \Psi_3 (V_{ab} W_{cd} + W_{ab} V_{cd}) + \Psi_4 V_{ab} V_{cd} \nonumber\,,
\end{align}
with the five complex coefficients defined by
\begin{align}
\nonumber
\Psi_0 &\equiv C_{abcd} k^a m^b k^c m^d \,, \\
\nonumber
\Psi_1 &\equiv C_{abcd} k^a l^b k^c m^d \,, \\
\nonumber
\Psi_2 &\equiv C_{abcd} k^a m^b \bar{m}^c l^d \,, \\
\nonumber
\Psi_3 &\equiv C_{abcd} k^a l^b \bar{m}^c l^d \,, \\
\label{PsiDef}
\Psi_4 &\equiv C_{abcd} \bar{m}^a l^b \bar{m}^c l^d \,. 
\end{align}
Therefore, determining the ten independent components of the Weyl tensor in (\ref{defweyl}) is equivalent to determining the five complex scalars defined above. With regards to their physical interpretation: $\Psi_0$ and $\Psi_1$ represent transverse and longitudinal waves in the $\mathbf{l}$ direction, $\Psi_2$ a Coulomb-like component and $\Psi_3$ and $\Psi_4$ are longitudinal and transverse wave components in the $\mathbf{k}$ direction.

Petrov's classification by Penrose \cite{PENROSE1960171} characterizes the Weyl tensor according to principal null directions $\mathbf{k}$ with the property
\begin{equation} \label{pnd}
k_{[e} C_{a]bc[d} k_{f]}k^b k^c = 0
\end{equation}
There can be at most four such null vectors (p.n.d.'s). If a space-time admits four distinct p.n.d.'s it is called algebraically general (type I), otherwise it is algebraically special.

If $\mathbf{k}$ is a member of the null tetrad then equation (\ref{pnd}) is equivalent to $\Psi_0 = 0$. We can rotate to an arbitrary complex null tetrad $(\mathbf{m'}, \mathbf{\bar{m'}}, \mathbf{l'}, \mathbf{k'})$, where the coefficient $\Psi_0$ undergoes the transformation:
\begin{equation}
\Psi_0 = \Psi_0'-4z \psi_1' +6z^2 \Psi_2'-4z^3 \Psi_3' +z^4 \Psi_4'\,,
\end{equation}
with $z$ a complex number. So we see that the determination of principal null directions is equivalent to solving the quartic equation for $z$:
\begin{equation}
\Psi_0'-4z \psi_1' +6z^2 \Psi_2'-4z^3 \Psi_3' +z^4 \Psi_4' = 0\,,
\end{equation}
showing that there can be indeed four (complex) roots to this equation, that do not need to be different. Depending on the amount and multiplicity of the p.n.d.'s we get the classification in Figure \ref{fig:PetrovMults}.

\bibliographystyle{utphys}

\bibliography{PetrovML}{}

\providecommand{\href}[2]{#2}\begingroup\raggedright\begin{thebibliography}{10}

\bibitem{Stephani:2003tm}
H.~Stephani, D.~Kramer, M.~A.~H. MacCallum, C.~Hoenselaers, and E.~Herlt,
  \href{http://dx.doi.org/10.1017/CBO9780511535185}{{\em {Exact solutions of
  Einstein's field equations}}}.
\newblock Cambridge Monographs on Mathematical Physics. Cambridge Univ. Press,
  Cambridge, 2003.

\bibitem{MacCallum:2018csx}
M.~A.~H. MacCallum, ``{Computer algebra in gravity research},''
  \href{http://dx.doi.org/10.1007/s41114-018-0015-6}{{\em Living Rev. Rel.}
  {\bfseries 21} no.~1, (2018) 6}.

\bibitem{Karlhede1980}
A.~Karlhede, ``A review of the geometrical equivalence of metrics in general
  relativity,'' \href{http://dx.doi.org/10.1007/BF00771861}{{\em Gen Relat
  Gravit} {\bfseries 12} (1980) 693–707}.

\bibitem{Petrov:1954}
A.~Z. Petrov, ``{Klassifikacya prostranstv opredelyayushchikh polya
  tyagoteniya},'' {\em Uch. Zapiski Kazan. Gos. Univ.} {\bfseries 114} (1954)
  55–69.

\bibitem{Segre1884}
C.~Segre, ``Sulla teoria e sulla classificazione delle omografie in uno spazio
  lineare ad uno numero qualunque di dimensioni,'' {\em Memorie della R.
  Accademia dei Lincei} {\bfseries 3a: 127} (1884) .

\bibitem{2000CQGra..17.2885P}
D.~{Pollney}, J.~E.~F. {Skea}, and R.~A. {d'Inverno}, ``{Classifying geometries
  in general relativity: III. Classification in practice},''
  \href{http://dx.doi.org/10.1088/0264-9381/17/15/304}{{\em Classical and
  Quantum Gravity} {\bfseries 17} no.~15, (Aug., 2000) 2885--2902}.

\bibitem{He:2017aed}
Y.-H. He, {\em {Deep-Learning the Landscape}}, 6, 2017.
\newblock \href{http://arxiv.org/abs/1706.02714}{{\ttfamily arXiv:1706.02714
  [hep-th]}}.
\newblock q.v. interview in {\it Science}, Vol 365, July, 2019.

\bibitem{He:2017set}
Y.-H. He, ``{Machine-learning the string landscape},''
  \href{http://dx.doi.org/10.1016/j.physletb.2017.10.024}{{\em Phys. Lett. B}
  {\bfseries 774} (2017) 564--568}.

\bibitem{Krefl:2017yox}
D.~Krefl and R.-K. Seong, ``{Machine Learning of Calabi-Yau Volumes},'' {\em
  Phys. Rev. D} {\bfseries 96} no.~6, (2017) 066014,
  \href{http://arxiv.org/abs/1706.03346}{{\ttfamily arXiv:1706.03346
  [hep-th]}}.

\bibitem{Carifio:2017bov}
J.~Carifio, J.~Halverson, D.~Krioukov, and B.~D. Nelson, ``{Machine Learning in
  the String Landscape},'' {\em JHEP} {\bfseries 09} (2017) 157,
  \href{http://arxiv.org/abs/1707.00655}{{\ttfamily arXiv:1707.00655
  [hep-th]}}.

\bibitem{Ruehle:2017mzq}
F.~Ruehle, ``{Evolving neural networks with genetic algorithms to study the
  String Landscape},'' {\em JHEP} {\bfseries 08} (2017) 038,
  \href{http://arxiv.org/abs/1706.07024}{{\ttfamily arXiv:1706.07024
  [hep-th]}}.

\bibitem{He:2019nzx}
Y.-H. He and M.~Kim, ``{Learning Algebraic Structures: Preliminary
  Investigations},'' \href{http://arxiv.org/abs/1905.02263}{{\ttfamily
  arXiv:1905.02263 [cs.LG]}}.

\bibitem{Alessandretti:2019jbs}
L.~Alessandretti, A.~Baronchelli, and Y.-H. He, ``{Machine Learning meets
  Number Theory: The Data Science of Birch-Swinnerton-Dyer},''
  \href{http://arxiv.org/abs/1911.02008}{{\ttfamily arXiv:1911.02008
  [math.NT]}}.

\bibitem{He:2020fdg}
Y.-H. He and S.-T. Yau, ``{Graph Laplacians, Riemannian Manifolds and their
  Machine-Learning},'' \href{http://arxiv.org/abs/2006.16619}{{\ttfamily
  arXiv:2006.16619 [math.CO]}}.

\bibitem{He:2021oav}
Y.-H. He, ``{Machine-Learning Mathematical Structures},''
  \href{http://arxiv.org/abs/2101.06317}{{\ttfamily arXiv:2101.06317 [cs.LG]}}.

\bibitem{davies2021advancing}
A.~Davies, P.~Veli{\v{c}}kovi{\'c}, L.~Buesing, S.~Blackwell, D.~Zheng,
  N.~Toma{\v{s}}ev, R.~Tanburn, P.~Battaglia, C.~Blundell, A.~Juh{\'a}sz, {\em
  et al.}, ``Advancing mathematics by guiding human intuition with ai,'' {\em
  Nature} {\bfseries 600} no.~7887, (2021) 70--74.

\bibitem{He:2018jtw}
Y.-H. He, \href{http://dx.doi.org/10.1007/978-3-030-77562-9}{{\em {The
  Calabi\textendash{}Yau Landscape: From Geometry, to Physics, to Machine
  Learning}}}.
\newblock Lecture Notes in Mathematics. 5, 2021.
\newblock \href{http://arxiv.org/abs/1812.02893}{{\ttfamily arXiv:1812.02893
  [hep-th]}}.

\bibitem{Ruehle:2020jrk}
F.~Ruehle, ``{Data science applications to string theory},''
  \href{http://dx.doi.org/10.1016/j.physrep.2019.09.005}{{\em Phys. Rept.}
  {\bfseries 839} (2020) 1--117}.

\bibitem{Krippendorf:2020gny}
S.~Krippendorf and M.~Syvaeri, ``{Detecting Symmetries with Neural Networks},''
  \href{http://arxiv.org/abs/2003.13679}{{\ttfamily arXiv:2003.13679
  [physics.comp-ph]}}.

\bibitem{Krishnan:2020sfg}
C.~Krishnan, V.~Mohan, and S.~Ray, ``{Machine Learning ${\cal N}=8, D=5$ Gauged
  Supergravity},'' \href{http://dx.doi.org/10.1002/prop.202000027}{{\em
  Fortsch. Phys.} {\bfseries 68} no.~5, (2020) 2000027},
  \href{http://arxiv.org/abs/2002.12927}{{\ttfamily arXiv:2002.12927
  [hep-th]}}.

\bibitem{Chen:2020dxg}
H.-Y. Chen, Y.-H. He, S.~Lal, and M.~Z. Zaz, ``{Machine Learning Etudes in
  Conformal Field Theories},''
  \href{http://arxiv.org/abs/2006.16114}{{\ttfamily arXiv:2006.16114
  [hep-th]}}.

\bibitem{Liu:2021azq}
Z.~Liu and M.~Tegmark, ``{Machine-learning hidden symmetries},''
  \href{http://arxiv.org/abs/2109.09721}{{\ttfamily arXiv:2109.09721 [cs.LG]}}.

\bibitem{Alexander:2021rch}
S.~Alexander, W.~J. Cunningham, J.~Lanier, L.~Smolin, S.~Stanojevic, M.~W.
  Toomey, and D.~Wecker, ``{The Autodidactic Universe},''
  \href{http://arxiv.org/abs/2104.03902}{{\ttfamily arXiv:2104.03902
  [hep-th]}}.

\bibitem{Altman:2021pyc}
R.~Altman, J.~Carifio, X.~Gao, and B.~Nelson, ``{Orientifold Calabi-Yau
  Threefolds with Divisor Involutions and String Landscape},''
  \href{http://arxiv.org/abs/2111.03078}{{\ttfamily arXiv:2111.03078
  [hep-th]}}.

\bibitem{Gao:2021xbs}
X.~Gao and H.~Zou, ``{Machine Learning to the Orientifold Calabi-Yau with
  String Vacua},'' \href{http://arxiv.org/abs/2112.04950}{{\ttfamily
  arXiv:2112.04950 [hep-th]}}.

\bibitem{dInverno1971CLASSIFICATIONOT}
R.~A. d'Inverno and R.~A. Russell-Clark, ``Classification of the harrison
  metrics.,'' {\em Journal of Mathematical Physics} {\bfseries 12} (1971)
  1258--1263.

\bibitem{LM1988}
F.~W. Letniowski and R.~G. McLenaghan, ``An improved algorithm for quartic
  equation classification and petrov classification,''
  \href{http://dx.doi.org/10.1007/BF00758122}{{\em Gen. Rel. Grav.} {\bfseries
  20} (1988) 463–483}.

\bibitem{1991GReGr..23.1023A}
J.~E. {Aman}, R.~A. {D'Inverno}, G.~C. {Joly}, and M.~A.~H. {MacCallum},
  ``{Quartic equations and classification of Riemann tensors in general
  relativity},'' \href{http://dx.doi.org/10.1007/BF00756865}{{\em General
  Relativity and Gravitation} {\bfseries 23} no.~9, (Sept., 1991) 1023--1055}.

\bibitem{Zakhary:2003}
E.~Zakhary, K.~Vu, and J.~Carminati, ``A new algorithm for the petrov
  classification of the weyl tensor,''
  \href{http://dx.doi.org/10.1023/A:1024497708100}{{\em General Relativity and
  Gravitation} {\bfseries 35} (07, 2003) 1223--1242}.

\bibitem{doi:10.1063/1.1724257}
E.~Newman and R.~Penrose, ``An approach to gravitational radiation by a method
  of spin coefficients,'' \href{http://dx.doi.org/10.1063/1.1724257}{{\em
  Journal of Mathematical Physics} {\bfseries 3} no.~3, (1962) 566--578}.

\bibitem{PENROSE1960171}
R.~Penrose, ``A spinor approach to general relativity,''
  \href{http://dx.doi.org/10.1016/0003-4916(60)90021-X}{{\em Annals of Physics}
  {\bfseries 10} no.~2, (1960) 171--201}.

\bibitem{Tanatarov:2012gf}
I.~V. Tanatarov and O.~B. Zaslavskii, ``{What happens to Petrov classification
  on horizons of axisymmetric dirty black holes},''
  \href{http://dx.doi.org/10.1063/1.4865995}{{\em J. Math. Phys.} {\bfseries
  55} (2014) 022502}, \href{http://arxiv.org/abs/1211.4376}{{\ttfamily
  arXiv:1211.4376 [gr-qc]}}.

\bibitem{Mathematica}
W.~R. Inc., ``Mathematica, {V}ersion 13.0.0.''
\newblock \url{https://www.wolfram.com/mathematica}. Champaign, IL, 2021.

\bibitem{kingma2017adam}
D.~P. Kingma and J.~Ba, ``Adam: A method for stochastic optimization,'' 2017.

\end{thebibliography}\endgroup

\end{document}